\documentclass[preprint,12pt]{article}

\usepackage[pdftex]{graphicx}
\usepackage{ccaption}
\captiondelim {.  } 

\usepackage[centertags]{amsmath}
\usepackage{amsthm,amsfonts,amssymb}
\usepackage{indentfirst}
\usepackage[font={scriptsize}]{caption}

\newcommand{\bfM}{\mathbf{M}}
\newcommand{\bfm}{\mathbf{m}}
\newcommand{\bfu}{\mathbf{u}}
\newcommand{\bfd}{\mathbf{d}}
\newcommand{\bfv}{\mathbf{v}}

\newcommand{\bfD}{\mathbf{D}}
\newcommand{\bfB}{\mathbf{B}}
\newcommand{\bfQ}{\mathbf{Q}}

\newtheorem{theorem}{Theorem}[section]

\title{Fitness landscape adaptation in open replicator systems with competition: application to cancer therapy}

\date{\vspace{-5ex}}

\author{ 	
	{\bf Igor Samokhin} \\
	Lomonosov Moscow State University\\ Moscow 119992, Russia\\
	{\bf Tatiana Yakushkina}\\ National Research University Higher School of Economics\\
	Moscow 101000, Russia\\
	{\bf Dmitry Markin}\\
	Lomonosov Moscow State University\\ Moscow 119992, Russia\\
	{\bf Alexander S. Bratus}\\
	Russian University of Transport\\Moscow 127994, Russia\\
	alexander.bratus@yandex.ru}

\begin{document}
	\maketitle

	\begin{abstract}
	This study focuses on open quasispecies systems with competition and death flow, described by modified Eigen and Crow-Kimura models. We examine the evolutionary adaptation process as a reaction to changes in rates. One of the fundamental assumptions, which forms the basis of our mathematical model, is the existence of two different timescales: internal dynamics time and evolutionary time. The latter is much slower and exhibits significant adaptation events. These conditions allow us to represent the whole evolutionary process through a series of steady-state equations, where all the elements continuously depend on the evolutionary parameter. The process can be seen as a formalization of Fisher's fundamental theorem of natural selection, with mean fitness adaptation.  Such models can describe systems of different cancer cell phenotypes affected by treatment, e.g., Chemotherapeutic agents.
	\end{abstract}

\section{Introduction}
\label{s1}
Adaptation to environmental changes is one of the fundamental concepts in evolutionary theory. 
Classical evolution systems like Crow-Kimura \cite{Crow1970}, Eigen \cite{Eigen1971}, and Schuster \cite{Shuster2011} models rely on the assumption of fixed population size 
at any time moment.  However, this approach can not be applied to the population with the direct elimination of one or several phenotypes. This scenario is essential for consideration 
in cancer and bacteria treatments, 
which target specific population types. In this paper, we call the targeted phenotypes the main types. In the previous studies \cite{Bratus2009, Egorov2020}, the mathematical apparatus for open replicator was developed using modified Eigen and Crow-Kimura models with explicit death flow. In the following subsection, we describe the model class and its modification for the research considered.
\subsection{Open quasispecies model with species competition}
Consider a population with genome length $l\in\mathbb{N}$. We suppose that each letter encode a gene in one of the two states: $0$ for wild one, and  $1$ for mutant. 
Suppose that each of the $n = 2^l$ types correspond to a phenotype. The distribution of types over the population is defined by the vector 
 $\bfu(t) = \left( u_1 \left( t \right), u_2 \left( t \right), \ldots, u_n \left( t \right) \right) \in \mathbb{R}_+ ^ n = [0, + \infty) ^ n$, which describes the numbers of sub-population at time moment $t$. The selection force is described through the fitness matrix  $\bfM = diag( m_1, m_2, \ldots, m_n)$, where elements $m_i$ stand for each type's replication rate. 
Mutation rates are given by a transition matrix $\bfQ = \left\{q_{ij} \right\}_{i,j = 1} ^ n$ with  $q_{ij}$ --- probability of the type $j$ being a result of $i$ replicating. This matrix satisfies the stochastic condition $\sum\limits_{i = 1} ^ n q_{ij} = 1, \quad j = \overline{1, n}$. In the model with explicit death flow, we introduced the death rates as elements of another matrix 
$\bfD = diag( d_1, d_2, \ldots, d_n)$. Lastly, $\bfB = ( \beta_{ij})_{i, j =1 } ^ n, \quad det|\bfB| \neq 0, \quad \beta_{ij} \geq 0, \quad i,j = \overline{1, n}$ is a competition matrix. 

Let $\chi(i,j)$ be the Hamming distance between genetic sequences $i$ and $j$. Suppose $q_{ij} = p ^ {\chi(i, j)} ( 1 - p) ^ {(l - \chi(i,j))}, \quad 0 < p \leq 1$. We call the following system of ODEs open quasispecies system \cite{Egorov2020}:
\begin{equation}
	\label{eq1.1}
	\begin{array}{l}
		\dfrac{d\bfu (t)}{dt} = \varphi \left( S \left( \bfu(t) \right) \right) \bfQ_M \bfu(t) - \bfD \bfu(t), \quad t > 0, \quad \bfu(0) = \bfu^0 \in \mathbb{R}_+^n, \\
		\bfQ_M = \bfQ\bfM, \quad S\left(\bfu(t) \right) = \sum\limits_{i = 1}^n \bfu_i (t).
	\end{array}
\end{equation}
Here,  $\varphi (S)$ is a smooth continuous function $S \in [0, +\infty)$ such as $S \varphi' ( S)$ is bounded over the set $S \in [0, +\infty)$. Without any loss of generality 
\begin{equation}
	\label{eq1.2}
	\varphi(S) = exp( -\gamma S), \quad \gamma > 0.
\end{equation}

The system \eqref{eq1.1} is positively invariant in $\mathbb{R}_+^ {n + 1}$ and the problem \eqref{eq1.1} has a unique solution for all $u^0 \in \mathbb{R}_+^ {n + 1}$ and  $t > 0$. Moreover, if $d_{min} = \min\limits_{1 \leq i \leq n} \{ d_i \} > 0$, then the total population size $S(\bfu)$ is bounded for any trajectory of the system \eqref{eq1.1}. The mean fitness of the system \eqref{eq1.1} can be found via the formula: 
\begin{equation*}
	f ( \bfu) = \left\{ 
	\begin{array}{l}
		0, \quad S( \bfu) = 0, \\
		\dfrac{ ( \bfm, \bfu) }{ ( \bfd, \bfu)}, \quad S ( \bfu) > 0.
	\end{array}
	\right.
\end{equation*}
In the latter system, we use round brackets $(\mathbf{a}, \mathbf{b})$  to define scalar product of the vectors  $ \mathbf{a}, \mathbf{b}$ in $\mathbb{R}^n$, $\mathbf{a} = (a_1, a_2, \ldots, a_n)$, $\mathbf{b} = (b_1, b_2, \ldots, b_n)$.

Hence, the mean fitness of the open system can be calculated as a fraction, where the nominator is the mean fitness of classical replicator system, and the denominator is the weighted death flow. One can note that  the previous open replicator models did not include competition explicitly, however, it plays a significant role in evolutionary dynamics.

If by $\left(\bfQ_M \bfu(t) \right)_i$, $\left( \bfD \bfu(t)\right)_i$ and $\left(\bfB \bfu (t) \right)_i$ we denote  $i$-th  component of the vector  $\bfQ_M \bfu(t)$, $\bfD\bfu(t)$ and $\bfB\bfu(t)$ respectively, then the modification of the Eigen open replicator system with competition can be written as follows:
\begin{equation}
	\label{eq1.3}
	\begin{array}{l}
		\dfrac{d u_i}{dt} (t) = \varphi \left( S (\bfu(t)) \right) \left( \bfQ_M \bfu(t)\right)_i - d_i u_i(i) - \left( \bfB\bfu(t) \right)_i u_i(t) , \\
		\quad u_i ( 0) = u_i^0, \quad i = \overline{1, n}, \quad u \in \mathbb{R}_+^ n, \\
		\bfQ_M = \bfQ\bfM, \quad S\left( \bfu(t) \right) = \sum\limits_{i = 1} ^ n u_i ( t).
	\end{array}
\end{equation}
The mean fitness of the system \eqref{eq1.3} can be calculated by the expression:
\begin{equation}
	\label{eq1.4}
	f (\bfu) = \left\{ 
	\begin{array}{l}
		0, \quad S(\bfu) = 0, \\
		\dfrac{ ( \bfm, \bfu) }{ (\bfd, \bfu) + ( \bfB\bfu, \bfu)}, \quad S (\bfu) > 0.
	\end{array}
	\right.
\end{equation}

\subsection{Open Crow-Kimura system with competition}

 The open permutation invariant Crow-Kimura model \cite{Crow1970} consist of $N + 1$ --- quantity of phenotypes classes, $\bfv(t) = \left(v_1(t), v_2(t), \ldots, v_{N + 1}(t)\right) \in \mathbb{R}_+ ^ {N + 1} = [0, + \infty) ^ {N + 1}$ is the population distribution  vector. $\bfM = diag(m_1, m_2, \ldots, m_{N + 1})$ fitness matrix with Malthusian coefficients as elements. $\mu$ mutation rate. $\mu > 0$. $\bfQ_N = \left\{q_{ij} \right\}_{i,j = 1} ^ {N + 1}$ 
three-diagonal transition matrix:  
\begin{equation*}
	\bfQ_N = \left(		\begin{tabular}{lllllllll}
		-N & 1 & 0 & 0 & \ldots & 0 & 0 & 0 & 0 \\
		N & -N & 2 & 0 & \ldots & 0 & 0 & 0 & 0 \\
		0 & N-1 & -N & 3 & \ldots & 0 & 0 & 0 & 0 \\
		\ldots & \ldots & \ldots & \ldots & \ldots & \ldots & \ldots & \ldots & \ldots \\
		0 & 0 & 0 & 0 & \ldots & 3 & -N & N-1 & 0 \\
		0 & 0 & 0 & 0 & \ldots & 0 & 2 & -N & N \\
		0 & 0 & 0 & 0 & \ldots & 0 & 0 & 1 & -N
	\end{tabular}
	\right).	
\end{equation*}
$\bfD = diag(d_1, d_2, \ldots, d_{N + 1})$ death rate matrix. $\bfB = ( \beta_{ij})_{i, j =1 } ^ {N + 1},$ ${\quad det|\bfB| \neq 0},$ $\quad \beta_{ij} \geq 0, \quad i,j = \overline{1, N + 1}$ competition matrix.

The following system describes the dynamics of the open Crow-Kimura model \cite{Egorov2020}:
\begin{eqnarray}
	\label{eq1.5}
	&&\dfrac{d\bfv (t)}{dt} = \varphi \left(S(\bfv(t)) \right) \left( \bfM + \mu \bfQ_N \right) \bfv(t) - \bfD \bfv(t),   \nonumber\\
	&&\bfv(0) = \bfv ^ 0 \in \mathbb{R}_+ ^ { N + 1},\quad S\left( \bfv(t) \right) = \sum\limits_{i = 1} ^ {N + 1} v_i ( t).
\end{eqnarray}
Here,  $\varphi (S)$ is a continuous smooth function  $S \in [0, +\infty)$ such as $S \varphi' (S)$ is bounded for $S \in [0, +\infty)$. Similarly to the previous section, without any loss of generality we take $\varphi( S)$ as \eqref{eq1.2}.

If $ \left( \left(\bfM + \mu \bfQ_N \right) \bfv(t) \right)_i$, $\left( \bfD \bfv (t) \right)_i$ and $\left( \bfB \bfv(t) \right)_i$  $i$-th component of the vector $\left(\bfM + \mu \bfQ_N \right) \bfv(t) $, $\bfD(t) $ and $\bfB\bfv(t)$ respectively,then the open system with competition has the form:
\begin{equation}
	\label{eq1.6}
	\begin{array}{l}
		\dfrac{ dv_i}{dt} (t) = \varphi \left( S \left( \bfv(t) \right) \right) \left( \left( \bfM + \mu \bfQ_N \right) \bfv(t)  \right)_i - d_i v_i(t) - \left( \bfB \bfv(t) \right)_i v_i(t), \\
		v_i (0) = v_i ^ 0, \quad i = \overline{ 1, N +1}, \quad v \in \mathbb{R}_+ ^ {N +1}, \quad S\left( \bfv(t) \right) = \sum\limits_{i = 0} ^ {N+1} v_i ( t).
	\end{array}
\end{equation}
The mean fitness of the system \eqref{eq1.6} is defined by the expression \eqref{eq1.4}
\begin{equation}
	\label{eq1.7}
	f (\bfv) = \left\{ 
	\begin{array}{l}
		0, \quad S(\bfv) = 0, \\
		\dfrac{ (\bfm, \bfv) }{ (\bfd, \bfv) + ( \bfB\bfv, \bfv)}, \quad S (\bfv) > 0.
	\end{array}
	\right.
\end{equation}
\section{Evolutionary adaptation problem. Methods and results}
\label{section2}

The systems \eqref{eq1.3}, \eqref{eq1.6} obtain qualitative differences from both classic Eigen and Crow-Kimura models and their open modifications \eqref{eq1.1}, \eqref{eq1.5} due to the explicit competition factor. In the previous settings, the steady-state solution are derived through the eigenvalue problems. In the case of \eqref{eq1.3} and \eqref{eq1.6}, the non-linear terms do not allow for the same approach. 

Studies of the population dynamics in bacteria \cite{bibid} has shown  that the therapeutic treatments can lead to resistant types, which are not susceptible to the drug used. 
Normally, the populations include one or several dominant phenotypes. In this study, we refer to such types as main ones. We assume that the main types exhibit both competitive and numerical advantage. If the direct effect of the treatment targets the main types, its death rate would significantly increase, changing their distribution among the population. This favors alternative types, which were originally less fit, to prosper in the new environment.
This study develops such a modification of microbiological evolutionary model, that describes the adaptation process to death rates changes and adjustment in population distribution towards new dominant types. Our key assumption of the proposed mathematical model is the following: the adaption process (as a reaction to direct treatment or elimination of particular types) happens in the slower evolutionary timescale and can not be described by the internal replicator dynamics. As in the previous studies \cite{Bratus2018, Bratus, Samokhin}, we introduce the iteration process to describe fitness landscape changes in evolutionary scale.

\paragraph{\ref{section2}.1}
Consider a variation of the quasispecies model, where the fitness landscape 
is changing over evolutionary time  $\tau = \varepsilon t$ with small enough coefficient $\varepsilon > 0$ for  internal dynamics time $t$. For the system \eqref{eq1.3}, the elements of the fitness matrix are smooth functions of the evolutionary parameter  $\tau$.
\begin{equation}
	\label{eq2.1}
	\bfM( \tau) = diag \left( m_1 (\tau), m_2 (\tau) , \ldots, m_n(\tau)\right), \quad \tau = \varepsilon t.
\end{equation} 

Taking into account the expression \eqref{eq2.1}, suppose that the distribution of types depends on  both the slow evolutionary time and the internal dynamics time. That is,  $\bfu = \bfu (t; \varepsilon t) = \left(u_1(t; \varepsilon t), u_2 ( t; \varepsilon t), \ldots, u_n ( t; \varepsilon t) \right)$. Hence, the dynamics of the open quasispecies system is described by the equations:
\begin{eqnarray*}
		\dfrac{d u_i}{dt} (t; \varepsilon t) =&& \varphi \left( S \left( \bfu \left( t; \varepsilon t \right) \right) \right) \left( \bfQ\bfM \left( \varepsilon t \right) \bfu \left( t; \varepsilon t \right) \right)_i - d_i u_i \left( t; \varepsilon t \right) -\\ &&\left( \bfB \bfu \left( t; \varepsilon t \right) \right)_i u_i \left( t; \varepsilon t \right), \\
		u_i ( 0; 0) = u_i^0, &&\quad i = \overline{1, n}, \bfu \in \mathbb{R}_+^ n, \quad S\left( \bfu ( t; \varepsilon t) \right) = \sum\limits_{i = 1} ^ n u_i ( t; \varepsilon t).
\end{eqnarray*}
If $t \leq T$, then for small enough  $\varepsilon > 0$ the behavior of the latter system if similar to the one for \eqref{eq1.3}.

Moving now to the dynamics of the system in evolutionary time. In this case, we have:
\begin{eqnarray*}
		\varepsilon \dfrac{d u_i}{dt} (\dfrac{\tau}{ \varepsilon }; \tau) =&& \varphi \left( S \left( \bfu \left( \dfrac{\tau}{ \varepsilon }; \tau \right) \right) \right) \left( \bfQ\bfM \left( \tau \right) \bfu \left( \dfrac{\tau}{ \varepsilon }; \tau \right) \right)_i - d_i u_i \left( \dfrac{\tau}{ \varepsilon }; \tau \right) - \\
		&&\left( \bfB \bfu \left( \dfrac{\tau}{ \varepsilon }; \tau \right) \right)_i u_i \left( \dfrac{\tau}{ \varepsilon }; \tau \right), \\
		u_i ( 0; 0) = u_i^0, &&\quad i = \overline{1, n},  u \in \mathbb{R}_+^ n, \quad S\left( \bfu \left( \dfrac{\tau}{ \varepsilon }; \tau \right) \right) = \sum\limits_{i = 1} ^ n u_i \left( \dfrac{\tau}{ \varepsilon }; \tau \right).
\end{eqnarray*}
For  $\varepsilon \rightarrow 0$, we derive:
\begin{equation}
	\label{eq2.2}
	\begin{array}{l}
		\varphi \left( S \left( \overline{\bfu} \left( \tau \right) \right) \right) \left( \bfQ\bfM \left( \tau \right) \overline{\bfu} \left( \tau \right) \right)_i - d_i \overline{u}_i \left( \tau \right) - \left( \bfB \overline{\bfu} \left( \tau \right) \right)_i \overline{u}_i \left( \tau \right) = 0, \\
		\overline{u}_i \left( \tau \right) = \lim\limits_{\varepsilon \rightarrow 0} u_i \left( \dfrac{ \tau}{ \varepsilon}; \tau \right), \quad i = \overline{1, n}, \quad \overline{\bfu} \in \mathbb{R}_+^ n, \quad S\left( \overline{\bfu} \left( \tau \right) \right) = \sum\limits_{i = 1} ^ n \overline{u}_i ( \tau).
	\end{array}
\end{equation}

Thus, the evolutionary adaptation process can be described by the steady-state equation, where the fitness landscape and distribution of types in the population depend on the evolutionary time $\tau$.

Another important assumption of the model is the maximization of the mean fitness as an adaptation instrument. We suppose that the theorem of natural selection can be understood 
in this sense: on the evolutionary timescale the system adapts to the death rate changes and the impact of the environment moving its fitness landscape towards its maximum mean value. 
According to the equation \eqref{eq1.4}, the mathematical definition of the mean fitness in the case of open quasispecies system is: 
\begin{equation}
	\label{eq2.3}
	f \left( \overline{ \bfu} \left( \tau \right) \right) = \left\{ 
	\begin{array}{l}
		0, \quad S \left( \overline{ \bfu} \left( \tau \right) \right) = 0, \\
		\dfrac{ \left( \bfm \left( \tau \right), \overline{\bfu} \left( \tau \right) \right) }{ (\bfd, \overline{\bfu} \left( \tau \right)) + ( B \overline{\bfu} \left( \tau \right), \overline{\bfu} \left( \tau \right))}, \quad S ( \overline{\bfu} \left( \tau \right)) > 0.
	\end{array}
	\right.
\end{equation}

To work with a well-posed maximization problem for the function \eqref{eq2.3}, we need to introduce  a limitation for the fitness landscape acceptance range. We assume  that  $\bfm \left( \tau \right) = \left( m_1 \left( \tau \right), m_2 \left( \tau \right), \ldots, m_n \left( \tau \right) \right)$ and
\begin{equation}
	\label{eq2.4}
	\bfm \left( \tau \right) \in \overline{ \mathbb{ M}}_n = \left\{ m_i \left( \tau \right) \geq \check{\bfm} > 0, \quad i = \overline{ 1, n}, \quad \sum\limits_{i =1} ^ n m_i \left( \tau \right) \leq K = const > 0 \right\}.
\end{equation}

For any fixed value of the evolutionary time  $\overline{ \tau}$, we can identify the internal dynamics of the system corresponding to the fitness landscape $\bfM \left( \overline{ \tau } \right)$:
\begin{equation*}
	\begin{array}{l}
		\dfrac{d u_i}{dt} (t; \overline{ \tau }) = \varphi \left( S \left( \bfu \left( t; \overline{ \tau } \right) \right) \right) \left( \bfQ\bfM \left( \overline{ \tau } \right) \bfu \left( t; \overline{ \tau } \right) \right)_i - d_i u_i \left( t; \overline{ \tau } \right) - \left( \bfB \bfu \left( t; \overline{ \tau } \right) \right)_i u_i \left( t; \overline{ \tau } \right) \\
		u_i ( 0; \overline{ \tau }) = u_i^0, \quad i = \overline{1, n}, \quad \bfu \in \mathbb{R}_+^ n, \quad S\left( \bfu ( t; \overline{ \tau }) \right) = \sum\limits_{i = 1} ^ n u_i ( t; \overline{ \tau }).
	\end{array}
\end{equation*}

\paragraph{\ref{section2}.2}
Similar derivations are valid for the Crow-Kimura model setting \eqref{eq1.6}.  Instead of the system \eqref{eq2.2}, in this case we have the following:
\begin{equation}
	\label{eq2.5}
	\begin{array}{l}
		\varphi \left( S \left( \overline{\bfv} \left( \tau \right) \right) \right) \left( \left( \bfM \left( \tau \right) + \mu \bfQ_N \right) \overline{\bfv} \left( \tau \right) \right)_i - d_i \overline{v}_i \left( \tau \right) - \left( \bfB \overline{\bfv} \left( \tau \right) \right)_i \overline{v}_i \left( \tau \right) = 0, \\
		\overline{v}_i \left( \tau \right) = \lim\limits_{\varepsilon \rightarrow 0} v_i \left( \dfrac{ \tau}{ \varepsilon}; \tau \right), \quad i = \overline{1, n}, \quad \overline{\bfv} \in \mathbb{R}_+^ {N + 1}, \quad S\left( \overline{\bfv} \left( \tau \right) \right) = \sum\limits_{i = 1} ^ {N + 1} \overline{v}_i ( \tau).
	\end{array}
\end{equation}
Analogously to the expression \eqref{eq2.3}, the mean fitness function is set by the formula:
\begin{equation}
	\label{eq2.6}
	f \left( \overline{ \bfv} \left( \tau \right) \right) = \left\{ 
	\begin{array}{l}
		0, \quad S \left( \overline{\bfv} \left( \tau \right) \right) = 0, \\
		\dfrac{ \left( \bfm \left( \tau \right), \overline{\bfv} \left( \tau \right) \right) }{ (\bfd, \overline{\bfv} \left( \tau \right)) + ( \bfB \overline{\bfv} \left( \tau \right), \overline{\bfv} \left( \tau \right))}, \quad S ( \overline{\bfv} \left( \tau \right)) > 0.
	\end{array}
	\right.
\end{equation}

To provide a well-posed maximization problem for  \eqref{eq2.6}, we introduce the limitation on the fitness landscape acceptance range. Here, we take the Malthusian coefficients as  $\bfm \left( \tau \right) = \left( m_1 \left( \tau \right), m_2 \left( \tau \right), \ldots, m_{N + 1} \left( \tau \right) \right)$ and
\begin{equation}
	\label{eq2.7}
	\bfm \left( \tau \right) \in \overline{ \mathbb{ M}}_{N + 1} = \left\{ m_i \left( \tau \right) \geq \check{\bfm} > 0,  i = \overline{ 1, N + 1},  \sum\limits_{i =1} ^ {N + 1} m_i \left( \tau \right) \leq K = const > 0 \right\}
\end{equation}

If we take a particular value $\overline{ \tau}$, we can derive a correspondent internal dynamics based on the fitness values $M \left( \overline{ \tau } \right)$:
\begin{eqnarray*}
		\dfrac{d v_i}{dt} (t; \overline{ \tau }) = &&\varphi \left( S \left( \bfv \left( t; \overline{ \tau } \right) \right) \right) \left( \left( \bfM \left( \overline{ \tau } \right) + \mu \bfQ_N \right) v \left( t; \overline{ \tau } \right) \right)_i - d_i v_i \left( t; \overline{ \tau } \right) -\\
		&& \left( \bfB \bfv \left( t; \overline{ \tau } \right) \right)_i v_i \left( t; \overline{ \tau } \right) \\
		v_i ( 0; \overline{ \tau }) = v_i^0, &&\quad i = \overline{1, N + 1},  v \in \mathbb{R}_+^ {N + 1}, \quad S\left( \bfv ( t; \overline{ \tau }) \right) = \sum\limits_{i = 1} ^ {N + 1} v_i ( t; \overline{ \tau }).
\end{eqnarray*}

Note that the proposed method is valid, if the systems \eqref{eq1.3} and \eqref{eq1.6} are permanent, i.e., there are unique equilibria $\overline{\bfu}$ for \eqref{eq1.3} and   $\overline{\bfv}$ for \eqref{eq1.6} respectively, such as:
\begin{equation*}
	\lim\limits_{t \rightarrow +\infty } \dfrac{ 1} { t} \int\limits_{0} ^ { t} \bfu \left( t \right)  dt = \overline{ \bfu} \in int \mathbb{ R}_+ ^ n, \quad \lim\limits_{t \rightarrow +\infty } \dfrac{ 1} { t} \int\limits_{0} ^ { t} \bfv \left( t \right)  dt = \overline{\bfv} \in int \mathbb{ R}_+ ^ {N + 1}.
\end{equation*}
\begin{theorem}
	\label{theorem1}
	Let the fitness landscape of the system \eqref{eq1.3} have positive values of Malthusian coefficients. Then, the system \eqref{eq1.3} is positively invariant  in  $\mathbb{R}_+ ^ n$, i.e., any trajectory that starts from  $\mathbb{R}_+ ^ {n}$  stay in this sat for $t > 0$.
	
	If $d_{min} = \min\limits_{1 \leq i \leq n} \{ d_i \} > 0$, then the solutions of the system \eqref{eq1.3}  with bounded positive evolutionary parameters are bounded  for $t \geq 0$.
\end{theorem}
Note that an equivalent statement holds for the Crow-Kimura setting of the system \eqref{eq1.6}.
We provide the discussing for the Theorem \ref{theorem1} in terms of system \eqref{eq1.3} , however, implying the same proof takes place for \eqref{eq1.6}.    

Consider the trajectories of the system \eqref{eq1.3}. If there is a equilibrium $\overline{ u} \in \partial \mathbb{R}_+ ^ n$ ($\overline{ v} \in \partial \mathbb{R}_+ ^ {N + 1}$) that 
is a saturated point \cite{Hofbauer2003}, then the trajectories could leave the area $\mathbb{R}_+ ^ n$ ($\mathbb{R}_+ ^ {N + 1}$). Let  $\overline{ \bfu} \neq 0$, then, with no loss of generality, $\overline{u}_k = 0$, $1 \leq k \leq n$. Hence,  $\overline{\bfu} = (\overline{ u}_1, \overline{ u}_2, \ldots, \overline{ u}_{k-1}, 0, \overline{ u}_{k+1}, \overline{ u}_{k+2}, \ldots, \overline{ u}_{n})$ and $\overline{u}_i \geq 0, \quad i = \overline{ 1, n}, \quad i \neq k, \quad \sum\limits_{ i =1, i \neq k} ^ {n} \overline{ u}_i > 0$. 

Define the function $V_k \left( t \right) = u_k( t)$ as: 
\begin{eqnarray*}
	\left. \dfrac{d V_k \left( t \right)}{dt} \right|_{ \bfu \left( t \right) = \overline{ \bfu}} = \left. \dfrac{d u_k \left( t \right)}{dt} \right|_{ \bfu \left( t \right) = \overline{ \bfu}} = \left. \varphi \left( S \left( \bfu \right) \right) \left( \bfQ_\bfM \bfu \right)_k \right|_{ \bfu = \overline{ \bfu}} =\\
	 \varphi \left( S \left( \overline{ \bfu} \right) \right) \sum\limits_{ j = 1, j \neq k} ^ {n} q_{kj} m_j \overline{ u}_j \geq 0.
\end{eqnarray*}
Therefore, the function $V_k( t) = u_k \left( t \right)$ is non-decreasing along the trajectories that start from $\overline{\bfu}$.
If $\overline{\bfu} = 0$, then the Jacobin of the system \eqref{eq1.3} at such point has the form:
\begin{equation*}
	\left. \dfrac{ \partial F_i \left(\bfu \right)}{ \partial u_j} \right|_{u = 0} = \left\{
	\begin{array}{l}
		q_{ii} m_i - d_i, \quad i = j, \\
		q_{ij} m_j, \quad i \neq j.
	\end{array}
	\right. , \quad i, j = \overline{ 1, n}. 
\end{equation*}
Since al non-diagonal elements of the Jacobian matrix are positive  $( q_{ij} m_j > 0, \quad i \neq j)$, then it does have a positive eigenvalue \cite{Bellman1960}. Hence, the equilibrium  $ \overline{\bfu} = 0$ is an unstable node.

Let us sum over \eqref{eq1.3}, using the identity:  
\begin{equation}
	\label{eq2.8}
	\sum\limits_{i, j = 1} ^ n q_{ij} m_j u_j \left( t \right) t = \sum\limits_{i = 1}^n m_i u_i \left( t \right).
\end{equation}
Then the following estimation is valid: 
\begin{eqnarray*}
	\dfrac{ d S ( \bfu)}{ dt} \leq m_{max} \varphi \left( S \left( \bfu \right) \right) S \left( \bfu \right) - d_{min} S \left( \bfu \right), \\ m_{max} = \max\limits_{1 \leq i \leq n} m_i, \quad d_{min} = \min\limits_{ 1 \leq i \leq n} d_i > 0.
\end{eqnarray*}
Since $\varphi(S) \leq \left( \gamma e \right) ^ {-1} = \theta > 0$, then
\begin{equation}
	\label{eq2.9}
	\dfrac{ dS}{dt} \leq m_{max} \theta - d_{min} S.
\end{equation}
Consider a one-dimensional dynamical system:
\begin{equation}
	\label{eq2.10}
	\dfrac{ d \xi(t)}{dt} = m_{max} \theta - d_{min} \xi( t), \quad \xi( t) \geq 0.
\end{equation}
Let us compare the dynamics of  \eqref{eq2.9} with  \eqref{eq2.10} with the same initial point $\xi(0) = S(0) > 0$. The system  \eqref{eq2.10} has a unique equilibrium 
\begin{equation*}
	\overline{\xi} = \dfrac{m_{max} \theta}{d_{min}} > 0,
\end{equation*}
which is an attractor.Therefore we get that  $S(t) \leq \xi(t)$. If $\xi(0) = S(0) > \overline{ \xi}$, then  $S(t) \leq \xi(t) \leq S(0)$. If otherwise $\xi(0) = S(0) < \overline{ \xi}$, then  $S(t) \leq \overline{\xi}$. Consequently, we derive:
\begin{equation*}
	S(t) \leq \max \left\{S\left(0\right), \overline{ \xi}\right\}.
\end{equation*}
Non-negativity of the solutions leads to its boundedness .

Consider similar steps to prove the theorem for the Crow-Kimura setting \eqref{eq1.6}. If $\overline{\bfv} \neq 0$, then, with no loss of generality, one can take $\overline{v}_k = 0$, where $1 \leq k \leq N + 1$. In this case, $\overline{ \bfv} = (\overline{ v}_1, \overline{ v}_2, \ldots, \overline{ v}_{k-1}, 0, \overline{ v}_{k+1}, \overline{ v}_{k+2}, \ldots, \overline{ v}_{N + 1}), \quad \overline{v}_i \geq 0, \quad i = \overline{ 1, N + 1}, \quad i \neq k, \quad \sum\limits_{ i =1, i \neq k} ^ {N + 1} \overline{ v}_i > 0$. We take $V_k \left( t \right) = v_k( t)$:
\begin{eqnarray*}
	\left. \dfrac{d V_{k} \left( t \right)}{dt} \right|_{ \bfv \left( t \right) = \overline{ \bfv}} = \left. \dfrac{d v_{k} \left( t \right)}{dt} \right|_{ \bfv \left( t \right) = \overline{ \bfv}} = \left. \varphi \left( S \left( \bfv \right) \right) \left( \bfQ_N \bfv \right)_k \right|_{ \bfv = \overline{ \bfv}} = \\
	\varphi \left( S \left( \overline{ \bfv} \right) \right) \sum\limits_{ j = 1, j \neq k} ^ {N+1} q_{k j} \overline{ v}_j \geq 0.
\end{eqnarray*}
Hence,  $V_k( t) = v_k \left( t \right)$ is non-decreasing along the trajectories with starting point $\overline{ v}$.
If  $\overline{ v} = 0$, then the Jacobian of the system \eqref{eq1.6} at this point is:
\begin{equation*}
	\left. \dfrac{ \partial F_i \left(\bfv \right)}{ \partial v_j} \right|_{\bfv = 0} = \left\{
	\begin{array}{l}
		m_i - \mu N - d_i, \quad i = j, \\
		\mu i, \quad i = j - 1, \\
		\mu \left( N + 1 - i \right), \quad i = j + 1.
	\end{array}
	\right. , \quad i, j = \overline{ 1, N + 1}, \quad \mu > 0. 
\end{equation*}
Within a positive factor $\overline{ \mu}$, this matrix coincides with the one for classical Crow-Kimura model. Hence, the equilibrium  $\overline{ v} = 0$ is an unstable node. 
To prove the boundedness of the solutions to \eqref{eq1.6}, we take the following expression: 
\begin{equation}
	\label{eq2.11}
	\sum\limits_{i = 1} ^ {N + 1} \left( Q_N u \right)_i = 0,
\end{equation}
which is valid due to the structure of the matrix  $\bfQ_N$. The rest of the derivations are the same as for \eqref{eq1.3}. $\square$

\section{Fitness variation with respect to evolutionary time $\tau$ changes}
\label{section3}

\paragraph{\ref{section3}.1}
Consider the system  \eqref{eq1.3}. Let $\overline{\bfu} \in int \mathbb{R}_+^n$ be an equilibrium of the system \eqref{eq1.3}, where $\tau \geq 0$ is the slow evolutionary time. The system \eqref{eq2.2} can be written as:
\begin{equation}
	\label{eq3.1}
	\begin{array}{l}
		\left( \bfQ_M \overline{ \bfu} \right)_i = \varphi ^ { -1} \left( S \left( \bfu \right) \right) \left[ \left( \bfD \overline{ \bfu} \right)_i + \left( \bfB \overline{ \bfu} \right)_i u_i \right], \quad i = \overline{ 1, n}, \\ 
		\bfQ_M \overline{ u} = \bfQ \bfM \left( \tau \right) \overline{ \bfu} \left( \tau \right), \quad \overline{ u}_i = \overline{ u}_i \left( \tau \right), \quad \tau \geq 0.
	\end{array}
\end{equation}
As we mentioned above, the explicit competition factor qualitatively changes the behavior of  the system \eqref{eq3.1}, which now can not be represented as a eigenvalue problem as was applicable to the classical open quasispecies system \eqref{eq1.1}.

Applying the equation \eqref{eq2.8} and summing up \eqref{eq3.1},we obtain: 
\begin{equation*}
	\varphi ^ {-1} \left( S \left( \overline{ \bfu} \left( \tau \right) \right) \right) = exp \left( \gamma S \right) \left( \overline{ \bfu} \left( \tau \right) \right) = \dfrac{ \left( \bfm, \overline{ \bfu} \left( \tau \right) \right)}{ \left( \bfd, \overline{ \bfu } \left( \tau \right) \right) + \left( \bfB \overline{ \bfu} \left( \tau \right), \overline{ \bfu} \left( \tau \right) \right)}.
\end{equation*}
From the mean fitness definition \eqref{eq2.3}, it follows that this expression provides the steady-state value of the mean fitness for  $\tau \geq 0$, i.e., 
\begin{equation}
	\label{eq3.2}
	f \left( \overline{ \bfu} \left( \tau \right) \right) = exp \left( \gamma S \left( \overline{ \bfu} \left( \tau \right) \right) \right).
\end{equation}
Hence, from \eqref{eq3.2},  we get:  if $\overline{ \bfu } \in int \mathbb{R}_+ ^ n$, then $f \left( \overline{ \bfu} \right) > 1$.

Consider the variation of the mean fitness value \eqref{eq3.2} over the evolutionary time change: from $\tau$ to $\tau + \Delta \tau$, $\Delta \tau > 0$. Here, we suppose the elements of the fitness landscape to be smooth functions of $\tau$.

Define as  $\delta \bfM, \delta \overline{\bfu}, \delta f ( \overline{ \bfu})$ the lead terms of the increments $\bfM( \tau)$, $\overline{\bfu}( \tau)$, and  $f( \overline{ u})$ respectively. According to Fisher's  theorem of natural selection, suppose that the mean fitness variation growths during the evolutionary process.

Let $\Delta \tau = \varepsilon$, $\varepsilon$  be a sufficiently small number. Then, within the order $\varepsilon ^ 2$, we derive from \eqref{eq3.1} the following equation:
\begin{eqnarray}
	\label{eq3.3}
		\left( \bfQ \delta \bfM \left( \tau \right) \overline{ \bfu} \right)_i + \left( \bfQ \bfM \left( \tau \right) \delta \overline{ \bfu} \left( \tau \right) \right)_i =\nonumber\\
		 f \left( \overline{ \bfu} \right) \left[ \bfD \delta \overline{ \bfu} \left( \tau \right) + \left( \left( \bfB + \bfB ^ T \right) \overline{ \bfu} \left( \tau \right) \right)_i \delta \overline{ u}_i \left( \tau \right) \right] +  \nonumber\\
		+ \delta f \left( \overline{ \bfu} \right) \left[ \left( \bfD \overline{ \bfu} \left( \tau \right) \right)_i + \left( \bfB \overline{ \bfu} \left( \tau \right) \right)_i \overline{ u}_i \left( \tau \right) \right].
\end{eqnarray}
Introduce $\Gamma \left( \overline{ u} \left( \tau \right) \right)$ as the diagonal matrix of the form:
\begin{equation}
	\label{eq3.4}
	\begin{array}{l}
		\Gamma \left( \overline{ \bfu} \left( \tau \right) \right) = diag \left( \left(\left( \bfB + \bfB ^ T \right) \overline{ \bfu} \left( \tau \right) \right)_1, \left(\left( \bfB + \bfB ^ T \right) \overline{\bfu} \left( \tau \right) \right)_2, \ldots, \left(\left( \bfB + \bfB ^ T \right) \overline{ \bfu} \left( \tau \right) \right)_n \right) = \\
		= diag \left( \sum\limits_{ j = 1} ^ n \left( b_{1j} + b_{j1} \right)  \overline{ u}_j \left( \tau \right), \sum\limits_{ j = 1} ^ n \left( b_{2j} + b_{j2} \right)  \overline{ u}_j \left( \tau \right), \ldots, \sum\limits_{ j = 1} ^ n \left( b_{nj} + b_{jn} \right)  \overline{ u}_j \left( \tau \right) \right) 
	\end{array}
\end{equation}
Define $w \left( \tau \right)$ as the vector: 
\begin{equation}
	\label{eq3.5}
	w( \tau ) = \left( w_{1}( \tau ), w_{2}( \tau ), \ldots, w_{n}( \tau ) \right), \quad w_{i}( \tau) = \left(\bf D \overline{ \bfu} \left( \tau \right)\right)_i + \left( \bfB \overline{ \bfu} \left( \tau \right) \right)_i \overline{ u}_i \left( \tau \right), \quad i = \overline{ 1, n}.
\end{equation}
Consider the matrix:
\begin{equation}
	\label{eq3.6}
	R \left( \tau \right) = \left[ \bfQ \bfM \left( \tau \right) - f \left( \overline{ \bfu} \left( \tau \right) \right) \left( \bfD + \Gamma \left( \overline{ \bfu} \left( \tau \right) \right) \right) \right] ^ { -1}.
\end{equation}

The increment $\delta \overline{ \bfu} ( \tau)$ can be derived from \eqref{eq3.3} using the expressions  \eqref{eq3.4}, \eqref{eq3.5} and \eqref{eq3.6}
\begin{equation}
	\label{eq3.7}
	\delta \overline{ \bfu} = R (\tau) \left( \delta f \left( \overline{ \bfu} \right) w \left( \tau \right) - \bfQ \delta \bfM \left( \tau \right) \overline{ \bfu} \left( \tau \right) \right).
\end{equation}

On the other hand, from the equation \eqref{eq3.2}, it follows that
\begin{equation}
	\label{eq3.8}
	\delta f \left( \overline{ \bfu} \right) = \gamma f \left( \overline{ \bfu} \right) \sum\limits_{ i =1 } ^ n \delta \overline{ u}_i \left( \tau \right) = \gamma f \left( \overline{ \bfu} \right) \left( \delta \overline{ \bfu} \left( \tau \right), I \right), \quad I = \left( 1, 1, \ldots, 1 \right) \in \mathbb{ R} ^ n.
\end{equation}

Then, from \eqref{eq3.7} and \eqref{eq3.8}, we can obtain the variation $\delta f \left( \overline{ u} \right)$:
\begin{equation}
	\label{eq3.9}
	\begin{array}{l}
		\delta f ( \overline{\bfu}) = \dfrac{ \left(R \left( \tau \right) \bfQ \delta \bfM \left( \tau \right) \overline{\bfu} \left( \tau \right), I \right)}{ \left( R \left( \tau \right) w \left( \tau \right), I \right) - \left( \gamma f \left( \overline{\bfu} \right) \right) ^ { -1}} = \dfrac{ \left( \delta \bfm \left( \tau \right), diag \left( \overline{ \bfu} \left( \tau \right) \right) \bfQ ^ T R ^ T \left( \tau \right) I \right) }{ \left( R \left( \tau \right) w \left( \tau \right), I \right) - \left( \gamma f \left( \overline{\bfu} \right) \right) ^ { -1}}, \\
		I = \left( 1, 1, \ldots, 1 \right) \in \mathbb{ R} ^ n, \quad \overline{ f } \left( \overline{ \bfu} \right) = exp \left( \gamma S \left( \overline{ \bfu} \right) \right), \quad \gamma > 0.
	\end{array}
\end{equation}
Since the variation of the fitness landscape matrix  coefficients can be  found in the nominator of the mean fitness formula \eqref{eq3.9}, then the variation of the fitness is a linear function of  $\delta m_i \left( \tau \right), \quad i = \overline{ 1, n}$. Taking  \eqref{eq2.4} into account, we get:
\begin{equation}
	\label{eq3.10}
	\sum\limits_{i = 1} ^ n \delta m_i \left( \tau \right) \leq 0, \quad \delta m_i \left( \tau \right) = \dfrac{ dm_i \left( \tau \right)}{ d \tau} \Delta \tau, \quad | m_i \left( \tau \right) | \leq k_i, \quad i = \overline{ 1 , n}.
\end{equation}
Based on the hypothesis about the slow changes of the fitness landscape, we have $k_i \approx \varepsilon$.

Let us write the condition for the local fitness maximum as the inequality:
\begin{equation*}
	\left( \delta \bfM \left( \tau \right), \bfQ ^ T R ^ T I \right) = \sum\limits_{ i = 1} ^ n \delta m_i \left( \tau \right) \overline{ u}_i \left( \tau \right) \left( \bfQ ^ T R ^ T I \right)_i \leq 0
\end{equation*}
over the set of variations $\delta m_i$ satisfying the condition \eqref{eq3.10}.

\paragraph{\ref{section3}.2}
Similar statements are valid for the Crow-Kimura model setting. Let $\overline{\bfv} \in int \mathbb{R}_+^{ N + 1}$  be the equilibrium of the system \eqref{eq1.6}, where $\tau \geq 0$  is slow evolutionary time. Summing the expressions  for $\left( \left( M \left( \tau \right) + \mu Q_N \right) \overline{ v} \left( \tau \right) \right)_i$ derived from \eqref{eq2.5} and using the equation \eqref{eq2.11}, we derive the formula for the mean fitness variation:
\begin{equation}
	\label{eq3.11}
	f \left( \overline{ \bfv} \left( \tau \right) \right) = exp \left( \gamma S \left( \overline{ \bfv} \left( \tau \right) \right) \right).
\end{equation}

The equation for variations \eqref{eq1.6} has the form:
\begin{eqnarray}
	\label{eq3.12}
		\left( \delta \bfM \left( \tau \right) \overline{ \bfv} \right)_i + \left( \left( \bfM\left( \tau \right) + \mu \bfQ_N \right) \delta \overline{ \bfv} \left( \tau \right) \right)_i = \nonumber\\f \left( \overline{ \bfv} \right) \left[ \bfD \delta \overline{ \bfv} \left( \tau \right) + \left( \left( \bfB + \bfB ^ T \right) \overline{ \bfv} \left( \tau \right) \right)_i \delta \overline{ v}_i \left( \tau \right) \right] + \nonumber \\
		+ \delta f \left( \overline{ \bfv} \right) \left[ \left( \bfD \overline{ \bfv} \left( \tau \right) \right)_i + \left( \bfB \overline{ \bfv} \left( \tau \right) \right)_i \overline{ v}_i \left( \tau \right) \right].
\end{eqnarray}
Define as  $\Gamma \left( \overline{ v} \left( \tau \right) \right)$ the diagonal matrix:
\begin{eqnarray}
	\label{eq3.13}
		\Gamma \left( \overline{ v} \left( \tau \right) \right) = diag \left( \left(\left( B + B ^ T \right) \overline{ v} \left( \tau \right) \right)_1,  \ldots, \left(\left( B + B ^ T \right) \overline{ v} \left( \tau \right) \right)_{N + 1} \right) = \nonumber\\
		= diag \left( \sum\limits_{ j = 1} ^ {N + 1} \left( b_{1j} + b_{j1} \right)  \overline{ v}_j \left( \tau \right),  \ldots, \sum\limits_{ j = 1} ^ {N + 1} \left( b_{N+1j} + b_{jN+1} \right)  \overline{ v}_j \left( \tau \right) \right). 
\end{eqnarray}
Let $w \left( \tau \right)$ denote the vector:
\begin{eqnarray}
	\label{eq3.14}
	w( \tau ) = \left( w_{1}( \tau ), w_{2}( \tau ), \ldots, w_{n}( \tau ) \right), \quad w_{i}( \tau) = \left( \bfD \overline{ \bfv} \left( \tau \right)\right)_i + \left( \bfB \overline{ \bfv} \left( \tau \right) \right)_i \overline{ v}_i \left( \tau \right), \\ i = \overline{ 1, N + 1}.\nonumber
\end{eqnarray}
Consider the matrix: 
\begin{equation}
	\label{eq3.15}
	R \left( \tau \right) = \left[ \bfM \left( \tau \right) + \mu \bfQ_N - f \left( \overline{ \bfv} \left( \tau \right) \right) \left( \bfD + \Gamma \left( \overline{\bfv} \left( \tau \right) \right) \right) \right] ^ { -1}.
\end{equation}

Express the variation $\delta \overline{ v} ( \tau)$ from \eqref{eq3.12}, using \eqref{eq3.13}, \eqref{eq3.14}, and \eqref{eq3.15}
\begin{equation}
	\label{eq3.16}
	\delta \overline{ v} = R (\tau) \left( \delta f \left( \overline{ u} \right) w \left( \tau \right) - \delta M \left( \tau \right) \overline{ v} \left( \tau \right) \right).
\end{equation}
From the other side,  \eqref{eq3.11} gives: 
\begin{equation}
	\label{eq3.17}
	\delta f \left( \overline{ \bfv} \right) = \gamma f \left( \overline{ \bfv} \right) \sum\limits_{ i =1 } ^ {N + 1} \delta \overline{ v}_i \left( \tau \right) = \gamma f \left( \overline{ \bfv} \right) \left( \delta \overline{ \bfv} \left( \tau \right), I \right), \quad I = \left( 1, 1, \ldots, 1 \right) \in \mathbb{ R} ^ {N + 1}.
\end{equation}

Hence, from  \eqref{eq3.16} and \eqref{eq3.17}, we can obtain $\delta f \left( \overline{ v} \right)$:
\begin{equation}
	\label{eq3.18}
	\begin{array}{l}
		\delta f ( \overline{\bfv}) = \dfrac{ \left(R \left( \tau \right) \delta \bfM \left( \tau \right) \overline{v} \left( \tau \right), I \right)}{ \left( R \left( \tau \right) w \left( \tau \right), I \right) - \left( \gamma f \left( \overline{\bfv} \right) \right) ^ { -1}} = \dfrac{ \left( \delta \bfm \left( \tau \right), diag \left( \overline{ \bfv} \left( \tau \right) \right) R ^ T \left( \tau \right) I \right) }{ \left( R \left( \tau \right) w \left( \tau \right), I \right) - \left( \gamma f \left( \overline{\bfv} \right) \right) ^ { -1}}, \\
		I = \left( 1, 1, \ldots, 1 \right) \in \mathbb{ R} ^ {N + 1}, \quad \overline{ f } \left( \overline{ \bfv} \right) = exp \left( \gamma S \left( \overline{ \bfv} \right) \right), \quad \gamma > 0.
	\end{array}
\end{equation}
Taking  \eqref{eq2.7} into account, we get: 
\begin{equation}
	\label{eq3.19}
	\sum\limits_{i = 1} ^ {N + 1} \delta m_i \left( \tau \right) \leq 0, \quad \delta m_i \left( \tau \right) = \dfrac{ dm_i \left( \tau \right)}{ d \tau} \Delta \tau, \quad | m_i \left( \tau \right) | \leq k_i, \quad i = \overline{ 1 , N + 1}.
\end{equation}

\section{Numerical modeling: results and discussion}
\label{section4}

\paragraph{\ref{section4}.1}
Consider the system \eqref{eq1.3} with the number of types $n = 16$. Each type is described by the binary string with the length  $l = 4$. The fitness landscape is set by the coefficients $\bfm^0 = $(6, 5, 5, 4, 5, 4, 4, 3, 5, 4, 4, 3, 4, 3, 3, 2). Mutation matrix $\bfQ$ is defined by transition coefficients $q_{ij} = p ^ {\chi(i, j)} ( 1 - p) ^ {(l - \chi(i,j))}, \quad p = 0.9, \quad i,j = \overline{1,16}$. Here,  $p$ stands for error-less replication rate, and $\chi(i,j)$  is the Hamming distance between $i$ and $j$. The function $\varphi \left( S \left( \bfu \right) \right) = exp \left( - \gamma \sum\limits_{ i =1 } ^ { 16} u_i \right),  \gamma = 1$. Competition matrix is given as $\bfB = \left\{ b_{ij} \right\}_{i,j = 1} ^ {16}$, $b_{ii} = 10 ^ { -4}$, $b_{ij} = 10 ^ { -5}, \quad i \neq j, \quad i,j = \overline{ 1 , 16}$. Death rates have the values $\overline{ d}^0 = $(0.0025, 0.0035, 0.0035, 0.005, 0.0035, 0.005, 0.005, 0.0071, 0.0035, 0.005, 0.005, 0.0071, 0.005, 0.0071, 0.0071, 0.01). The set \eqref{eq2.4} is introduced by  $\check{m} = 2, \quad K = 64$. For the simulation process, we take the evolutionary time step as $\Delta \tau = 10 ^ { -3}$. 

The fist type has both numerical and competitive advantage: $m^0_{1} = \max\limits_{i = \overline{ 1, 16}} m^0_{i} = 6$,  $d^0_{1} = \min\limits_{ i = \overline{ 1, 16}} d^0_{i} = 0.0025$. 

According to our numerical analysis, the fitness landscape  adaptation process during the mean fitness maximization with the parameter  $\Delta \tau$ was completed  after 28002 iterations  of algorithm described in \ref{section3}. At the initial time,  the steady-state distribution of the population has the value $\overline{\bfu}_0=$(3.2777, 0.7526, 0.7526, 0.1504, 0.7526, 0.1504, 0.1504, 0.0282, 0.7526, 0.1504, 0.1504, 0.0282, 0.1504, 0.0282, 0.0282, 0.0051). The mean fitness is calculated: $ln \left( f \left( \overline{ u}_0 \right) \right) = 7.308$. After the end of evolutionary process, the first type gained absolute advantage in terms of the fitness coefficients $m^1$, i.e.,  $m^1_{1} = 34$. At the same time, the other two types obtained the minimal possible fitness values over acceptance range: $m^1_{i} = 2, \quad i = \overline{ 2, 16}$. The numbers of the types at the end of evolutionary process turned to $\overline{ u}_{1} = $(6.0976, 0.6087, 0.6087, 0.0533, 0.6087, 0.0533, 0.0533, 0.0048, 0.6087, 0.0533, 0.0533, 0.0048, 0.0533, 0.0048, 0.0048, 0.0005). For such new distribution, the mean fitness reached the value $ln \left( f \left( \overline{ u}_{1} \right) \right) = 8.872$, i.e., increased 4.778 times.

\begin{figure}[htp]
	\begin{center}
		\includegraphics[width = 0.6\textwidth]{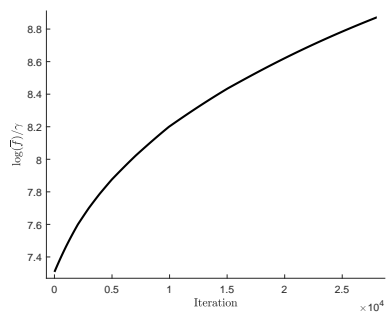}
		\caption{Case 4.1.The mean fitness growth compared to the number of iterations of the evolutionary adaptation algorithm.}
		\label{figure:system_base_f_MUTfitness_ex1}
	\end{center}
\end{figure}

\begin{figure}[htp]
	\begin{center}
		\includegraphics[width = 0.6\textwidth]{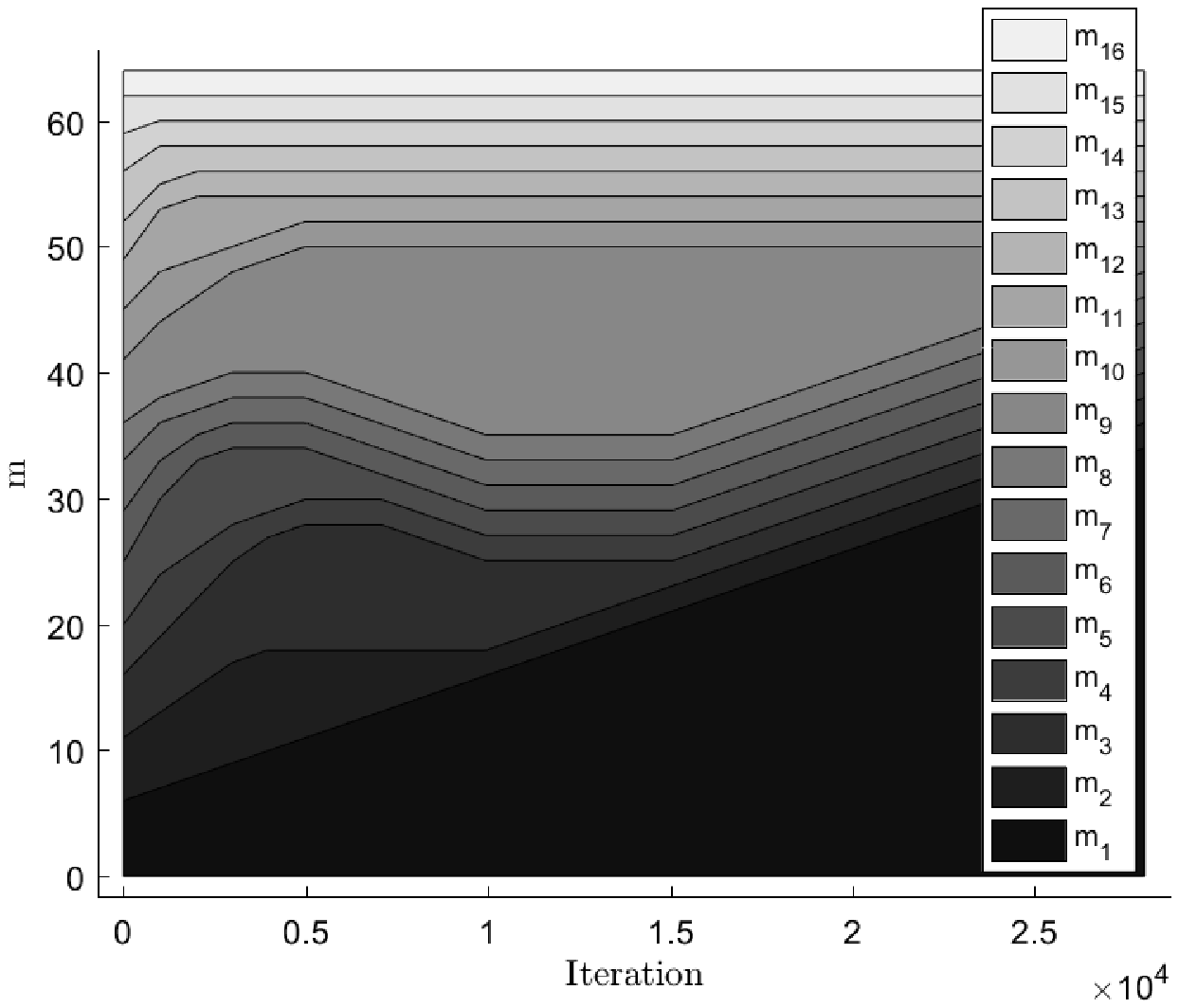}
		\caption{Case 4.1. The fitness landscape at steady-state compared to the number of iterations of the evolutionary adaptation algorithm.}
		\label{figure:system_base_f_MUTm_cumsum_ex1}
	\end{center}
\end{figure}

The mean fitness growth for  $f \left( \overline{ \bfu} \left( \tau \right) \right)$ depending on the number of steps in the adaptation process in timescale $\tau$ is shown at  Fig.~\ref{figure:system_base_f_MUTfitness_ex1}. The values of the fitness landscape at the steady-state depending on the number of steps in the adaptation process are shown at Fig.~\ref{figure:system_base_f_MUTm_cumsum_ex1}. 

\paragraph{\ref{section4}.2}
Consider the situation, where the death rate of the first type significantly increased at the very end of the adaptation process. It can be interpreted as the impact of the therapeutic agent introduced to the system. Moreover, suppose it has the impact, proportional to the Hamming distance to the first type. 

The updated death rate vector has the form $\overline{d}^2 = $(0.2, 0.095, 0.095, 0.045, 0.095, 0.045, 0.045, 0.021, 0.095, 0.045, 0.045, 0.021, 0.045, 0.021, 0.021, 0.01). The fitness landscape  $m^2$ corresponds to the final value of $m^1$ from the previous evolutionary process. All other parameters remained the same, as at the previous stage. The numbers of the types and the mean fitness of the system increased under the influence of therapy.

The fitness maximization process for this scenario required 32002 iterations of the adaptation process to complete the change in fitness landscape. At the start, we have the steady-state distribution $\overline{ u}_2=$(0.0020, 0.0038, 0.0038, 0.0341, 0.0038, 0.0341, 0.0341, 0.3410, 0.0038, 0.0341, 0.0341, 0.3410, 0.0341, 0.3410, 0.3410, 3.3024). The initial mean fitness is $ln \left( f \left( \overline{ u}_2 \right) \right) = 4.888$.  After the whole adaptation process, the sixteenth type became dominant in the fitness landscape $m^3$, i.e.,  $m^3_{16} = 34$, where all other values moved to its minimal border $m^3_{i} = 2, \quad i = 1, \overline{ 3, 16}$. The population distribution turned to the state $\overline{ u}_{3} = $(0.0001, 0.0011, 0.0011, 0.0193, 0.0011, 0.0193, 0.0193, 0.3508, 0.0011, 0.0193, 0.0193, 0.3508, 0.0193, 0.3508, 0.3508, 6.1274). The new fitness value increased 15.847 times compare to the previous stage: $ln \left( f \left( \overline{ u}_{3} \right) \right) = 7.651$.

\begin{figure}[htp]
	\begin{center}
		\includegraphics[width = 0.6\textwidth]{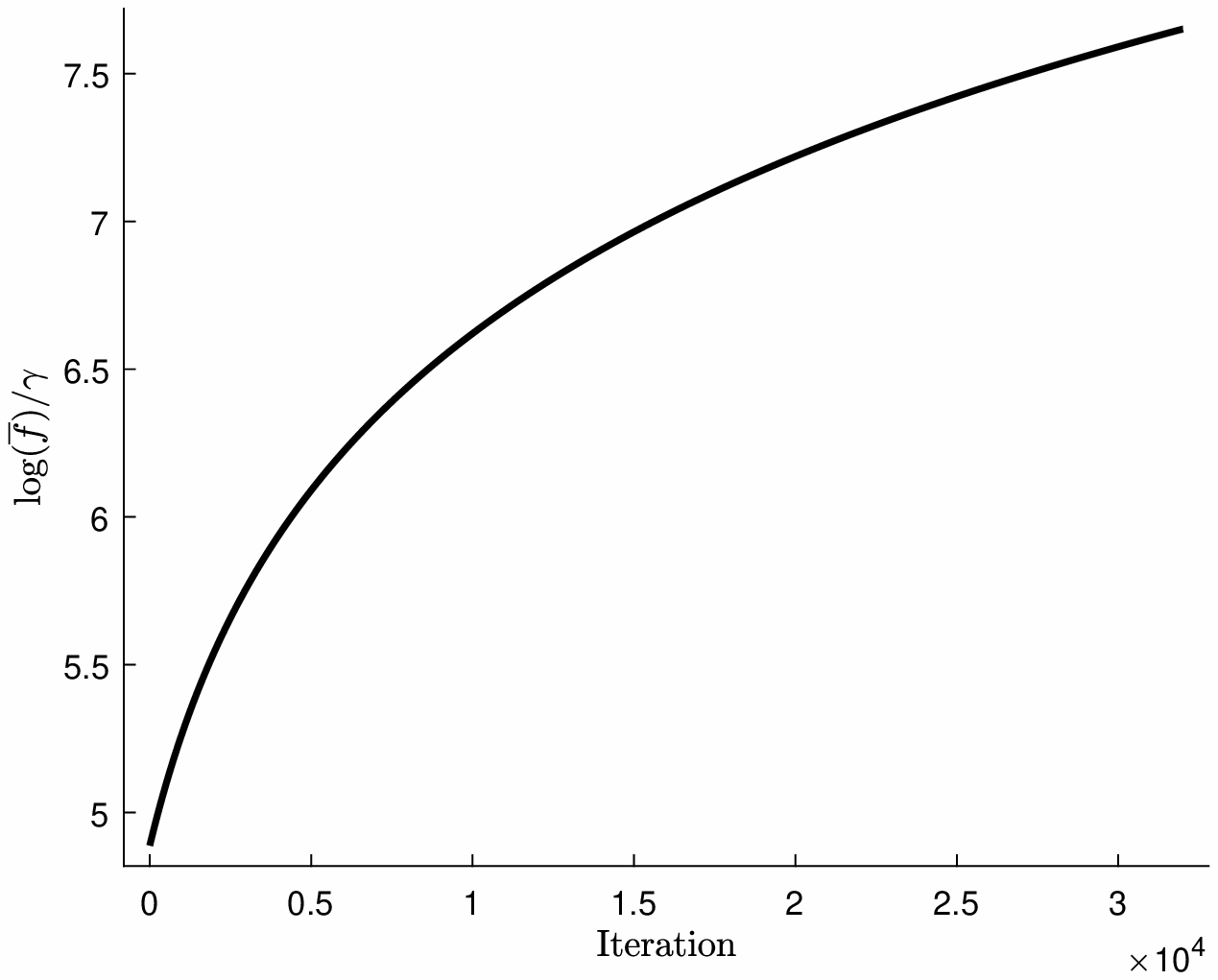}
		\caption{Case 4.2. Mean fitness increase compare to the number of steps in the adaptation process}
		\label{figure:system_base_f_MUTfitness_ex2}
	\end{center}
\end{figure}

\begin{figure}[htp]
	\begin{center}
		\includegraphics[width = 0.6\textwidth]{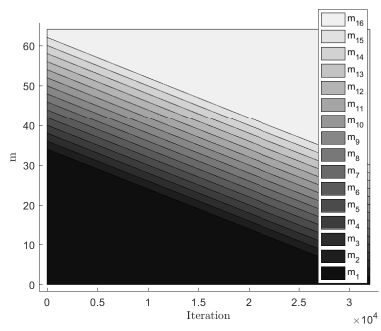}
		\caption{ Case 4.2. Fitness landscape at steady-state compare to the number of steps in the iteration process}
		\label{figure:system_base_f_MUTm_cumsum_ex2}
	\end{center}
\end{figure}

The mean fitness dynamics $f \left( \overline{ u} \left( \tau \right) \right)$ depending on the number of iterations in the adaptation algorithm is shown at Fig. \ref{figure:system_base_f_MUTfitness_ex2}. In Fig. \ref{figure:system_base_f_MUTm_cumsum_ex2}, we present the fitness landscape values at the steady-state depending on the number of steps in the iteration process.  

Therefore, we showed how the first type lost its evolutionary advantage and other types became dominant, depending in their better resilience to the therapeutic agent.

\paragraph{\ref{section4}.3}
Consider a different case: the system \eqref{eq1.6} with $N + 1 = 9$ types. The  mutation rate is  $\mu = 1$. The fitness landscape is set as $m^0 = $(4, 4, 4, 8, 8, 8, 4, 4, 4). Introduce the function  $\varphi \left( S \left( \bfv \right) \right) = exp \left( - \gamma \sum\limits_{ i =1 } ^ { 9} v_i \right), \quad \gamma = 0.1$. The competition matrix is  $\bfB = \left\{ b_{ij} \right\}_{i,j = 1} ^ {9}$, where 
\begin{equation*}
	b_{ij} = \left\{ 
	\begin{array}{l}
		10 ^ { -4}, \quad i = j; \\
		10 ^ { -5}, \quad i > j; \\
		10 ^ { -6}, \quad i < j.
	\end{array}
	\right.
\end{equation*}
The death rates have the values $\overline{ d}^0 = $(0.003, 0.003, 0.003, 0.001, 0.001, 0.001, 0.003, 0.003, 0.003). The set \eqref{eq2.7} is described by $\check{m} = 3, \quad K = 48$. For modeling, we take the step for the evolutionary timescale $\Delta \tau = 10 ^ { -3}$. 

Here, the types with numbers 3, 4, 5 exhibit evolutionary advantage having both better fitness and competitive traits. 

The complete adaptation process takes   16002 iterations of the algorithm with the step $\Delta \tau$  and logic described in \ref{section3}. At the start, we have the population distribution  $\overline{ v}_0=$(0.1164, 1.2126, 5.9915, 20.4500, 24.6832, 18.2542, 4.7758, 0.8446, 0.0713). The initial mean fitness value is  $ln \left( f \left( \overline{ u}_0 \right) \right) = 7.640$. After the end of the adaptation process, the fifth type gained the advantage in the fitness landscape $m^1$, i.e.,  $m^1_{5} = 24$. At the same time, all other types have the lowest possible fitness $m^1_{i} = 3, \quad i = 1, 2, \overline{4, 16}$. The population distribution transformed to $\overline{ v}_{1} = $(0.0270, 0.4077, 2.9912, 15.0544, 46.3471, 13.4717, 2.3346, 0.2735, 0.0157). The mean fitness increased 1.571 times: $ln \left( f \left( \overline{ u}_{1} \right) \right) = 8.092$.
\par
\begin{figure}[htp]
	\begin{center}
		\includegraphics[width = 0.6\textwidth]{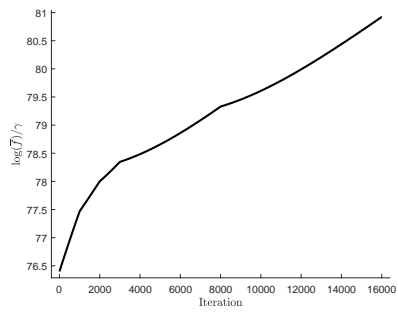}
		\caption{Case 4.3. The mean fitness values depending on the number of steps in adaptation process}
		\label{figure:system_base_f_MUTfitness_ex3}
	\end{center}
\end{figure}

\begin{figure}[htp]
	\begin{center}
		\includegraphics[width = 0.6\textwidth]{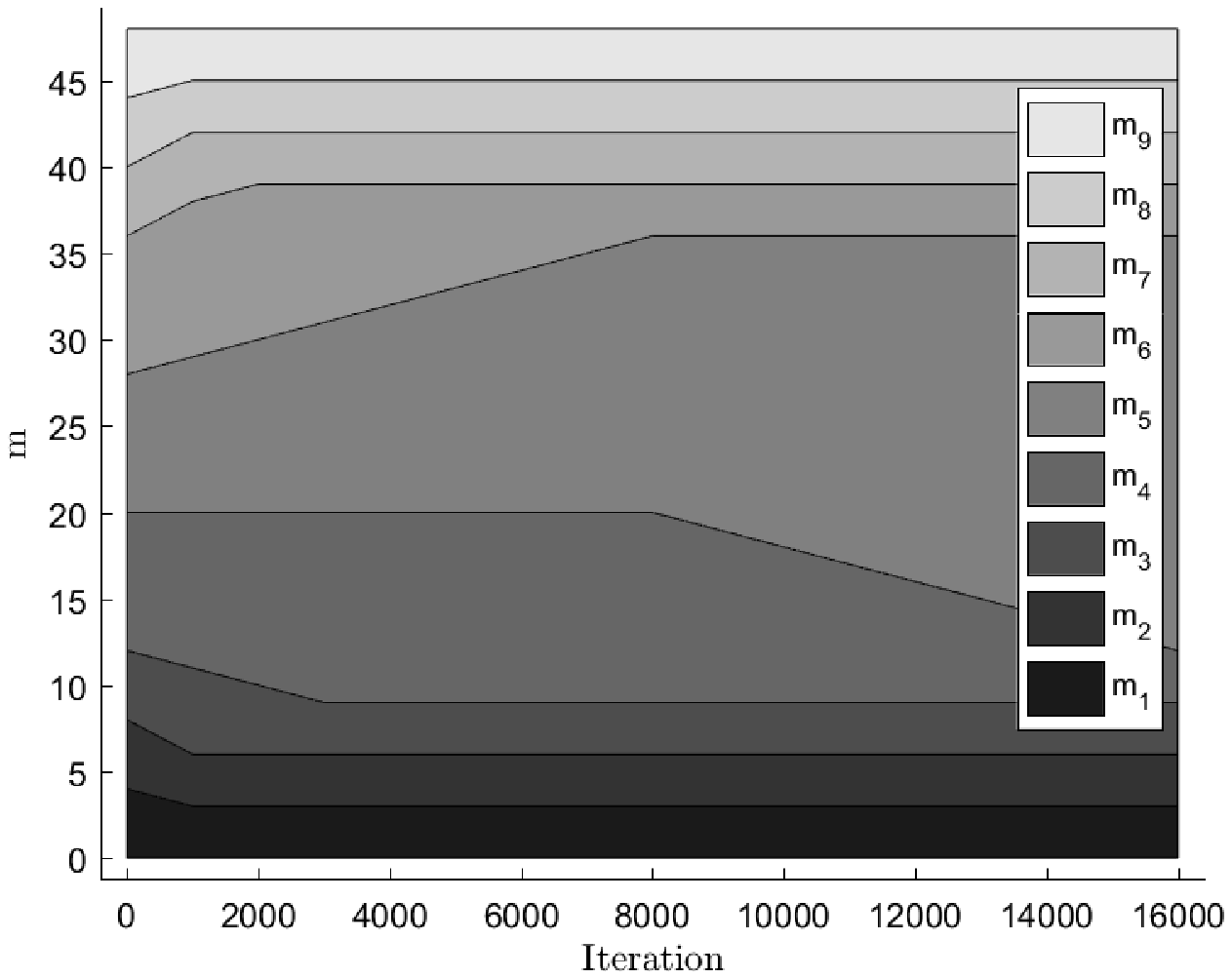}
		\caption{Case 4.3. The mean fitness values in the steady-state depending on the number of step in adaptation process}
		\label{figure:system_base_f_MUTm_cumsum_ex3}
	\end{center}
\end{figure}

 The mean fitness growth dynamics is shown $f \left( \overline{ v} \left( \tau \right) \right)$ in timescale $\tau$ Fig. \ref{figure:system_base_f_MUTfitness_ex3}. The fitness landscape at the steady-state changing over the iteration process is presented in Fig. \ref{figure:system_base_f_MUTm_cumsum_ex3}. 
\paragraph{\ref{section4}.4}
Let the death rates of types 4, 5, 6  significantly increase at the end of the evolutionary process, while all other types' death rates remain unchanged. In this case, the updated vector takes the form $\overline{d}^2 = $(0.003, 0.003, 0.003, 0.3, 0.3, 0.3, 0.003, 0.003, 0.003). The fitness landscape $m^2$ corresponds to the value $m^1$ from the previous evolutionary step. The rest of the model parameters coincide with the ones in the previous example.
The fitness landscape  change during maximization with the parameter $\Delta \tau$ completed after  21002 iteration of the algorithm, described in \ref{section3}. At the initial moment, we have $\overline{ u}_2=$(0.7795, 4.0534, 7.4645, 3.5935, 10.6241, 3.4945, 7.0681, 3.8110, 0.7296). Here, the logarithm of the fitness is $ln \left( f \left( \overline{ u}_2 \right) \right) = 4.162$.The second type eventually dominated in the landscape $m^3$, i.e., $m^3_{2} = 24$, where all the other fitness coefficients reached their admissible minimum $m^3_{i} = 3, \quad i = 1, \overline{ 3, 16}$. The population distribution became $\overline{ u}_{3} = $(4.1241, 51.2710, 21.0588, 0.1955, 0.0015, 0.0000, 0.0000, 0.0000, 0.0000), where $ln \left( f \left( \overline{ u}_{3} \right) \right) = 7.665$. This means the increase in the mean fitness by 33.215 times.
\par
\begin{figure}[htp]
	\begin{center}
		\includegraphics[width = 0.6\textwidth]{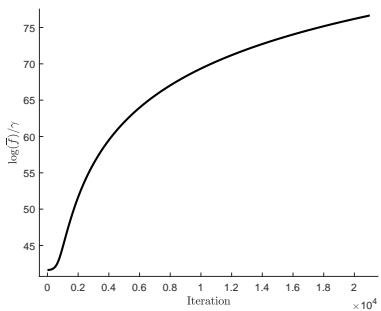}
		\caption{Case 4.4. Mean fitness values compare to the number of steps in the adaptation process}
		\label{figure:system_base_f_MUTfitness_ex4}
	\end{center}
\end{figure}

\begin{figure}[htp]
	\begin{center}
		\includegraphics[width = 0.6\textwidth]{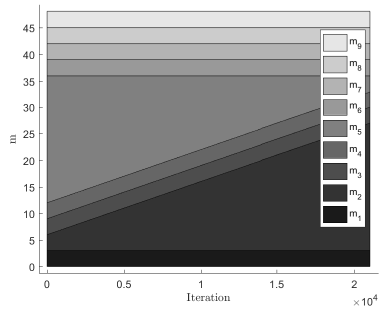}
		\caption{Case 4.4. Fitness landscape values at steady-state compare to the number of steps in the adaptation process}
		\label{figure:system_base_f_MUTm_cumsum_ex4}
	\end{center}
\end{figure}

 The change of the fitness function $f \left( \overline{ v} \left( \tau \right) \right)$ over evolutionary time is given in Fig. \ref{figure:system_base_f_MUTfitness_ex4}, where the steady state is described by Fig. \ref{figure:system_base_f_MUTm_cumsum_ex4}. 
%
\section{Conclusion}
In this study, we proposed such modifications of mathematical models for microbiological evolution that describe the adaptation process to the environmental changes, in particular, to the increased death rates for some of the types in the population. These conditions were inspired by real examples of therapeutic practices for cancer cells and bacteria, where specific phenotypes are eliminated. It is known \cite{Fox2010, Loeb2011} that in this case the populations can  develop therapy-resistant phenotypes, which the evolutionary advantage. In the previous study of our research group \cite{Samokhin}, we showed that the changes in fitness landscape play the leading role in the adaptation process. Here, we develop the idea further and consider the impact of the competition between different types. Suggested models include the explicit death flow and competition influence, showing significant changes of the population distribution. The numerical simulation demonstrated how new types can become dominant as a result of the adaptation process, which coincide with increased fitness coefficients and changed fitness landscape. 

\section*{Acknowledgedments}
The work is supported by the Russian Science Foundation under grant 19-11-00008.

\end{document}